\documentclass[useAMS,usenatbib]{mn2e}
\usepackage{times}
\usepackage{amssymb}
\usepackage{epsfig}
\newcommand{\rmd}{\mathrm{d}}
\begin{document}
\title{On de-Sitter Geometry in Cosmic Void Statistics}
\author[G.~W.~Gibbons, M.~C.~Werner, N.~Yoshida and S. Chon]{G.~W.~Gibbons$^1$, M.~C.~Werner$^{2,4}$\thanks{E-mail:marcus.werner@ipmu.jp}, N.~Yoshida$^{2,3}$ and S.~Chon$^3$\\ 
$^1$Department of Applied Mathematics and Theoretical Physics, University of Cambridge,
Wilberforce Road, Cambridge, CB3 0WA, UK\\
$^2$Kavli Institute for the Physics and Mathematics of the Universe (WPI), University of
Tokyo, 5-1-5 Kashiwanoha, Kashiwa, 277-8583, Japan, \\
$^3$Department of Physics, University of Tokyo, 7-3-1 Hongo, Tokyo, 113-0033, Japan, \\
$^4$Department of Mathematics, Duke University, Durham, NC 27708, USA}
\date{This draft \today}
\maketitle

\pagerange{\pageref{firstpage}--\pageref{lastpage}} \pubyear{0000}
\label{firstpage}
\begin{abstract}
Starting from the geometrical concept of a 4-dimensional de-Sitter configuration of spheres in Euclidean 3-space and modelling voids in the Universe as spheres, we show that a uniform distribution over this configuration space implies a power-law for the void number density which is consistent with results from the excursion set formalism and with data, for an intermediate range of void volumes. The scaling dimension of the large scale structure can be estimated as well. We also discuss the effect of restricting the survey geometry on the void statistics. This work is a new application of de-Sitter geometry to cosmology and also provides a new geometrical perspective on self-similarity in cosmology.
\end{abstract}

\section{Introduction}
While the existence and characteristic distributions of empty regions are already implicit in early self-similar models of structure hierarchy in the Universe (cf. \cite{Ma82}, \cite{Jo-etal04}), it was the observational discovery of large, approximately spherical regions almost devoid of visible galaxies known as voids and supervoids (e.g., \cite{Ei-etal80}, \cite{Ki-etal81}) that sparked further theoretical work on underdense regions in the Universe. This included analytical studies of the dynamical evolution of individual voids and shell-crossing (\cite{Pe82,Sa82}) as well as statistical properties of the void distribution. Applying a simple sign-reversal argument, \cite{Ic84} pointed out that the sphericity of underdense regions increases due to gravitational dynamics, so that it is in fact natural to expect approximately spherical voids. Early theoretical approaches to void statistics used Poisson statistics of empty regions, as developed by \cite{PoPr86}, Voronoi tesselations (\cite{IcvdWe87}), and structure formation theory in analogy to the statistics of overdensity peaks (\cite{Ba-etal86}) in Gaussian primordial density fluctuations (e.g., \cite{Be90}). Self-similar features in the void size distribution were also noted (e.g., \cite{Ei-etal89}). Of course, much of this earlier work concentrated on Einstein-de-Sitter cosmology with $\Omega_M=1$, as was favoured then. Especially after the discovery of the near-isotropic CMB, it was realized that the largest observable voids were difficult to explain theoretically and may therefore provide important cosmological constraints (e.g., \cite{Bl-etal92}). Attention was also drawn to galaxy properties within voids (e.g., the void phenomenon of \cite{Pe01}). Now, with the advent of the $\Lambda$CDM paradigm, detailed numerical studies of void statistics (e.g., \cite{Co-etal05} for a comparison) and analytical studies of void shape evolution in redshift space (e.g., \cite{MaSaTr11}) have been conducted. Moreover, the void formation theory based on the excursion set formalism of primordial density fluctuations has been refined to include the notion of hierarchy (\cite{ShvdWe04}). This formalism has been extended more recently to investigate effects of non-Gaussianity (e.g., \cite{dAm-etal11}) and modified theories of gravity (e.g., \cite{ClCaLi13}) on voids. The possibility of using voids for precision cosmology has also been explored (e.g., \cite{LaWa10}). On the observational side, a void catalogue extracted from the SDSS Data Release 7 has recently been published (\cite{Pa-etal12}), which illustrates the strong dependence of void statistics on the underlying void definition. The effect of survey masks and boundaries has been considered in detail by \cite{Su-etal12}, and void sphericity properties have been noted by \cite{TaVaMo13}. \cite{Hi-etal13} have shown how weak lensing observations may be used to measure the mass distribution of voids.
\newline
In this paper, we propose a new geometrical approach to void statistics based on the idea that voids can be modelled well by spheres in the Euclidean 3-space. Each such sphere is represented by a point in the 4-dimensional configuration space of spheres, with radius and centre position as coordinates. Since the spatial contact between these spheres is physically important (e.g., in the evolution of voids), it is interesting to consider transformations of the configuration space that preserve such contact relations, namely conformal transformations, in order to define a distance measure on the configuration space. Given these assumptions, it turns out that the configuration space has a de-Sitter geometry, with timelike radii and spacelike centre positions of the spheres, which emerges from classical sphere geometry and is different from cosmological de-Sitter spacetime. One can now study the size distribution of spherical voids in the Universe in terms of their distribution over this de-Sitter configuration space. The approach presented here has been inspired by other notions of configuration space measures in cosmology (e.g., \cite{GiTu08}), and another application of this de-Sitter configuration space of spheres in a different astronomical context has been proposed recently (\cite{GiWe13}).
\newline
The outline of this work is as follows. In Section \ref{sec:desitter}, we give a brief review of the de-Sitter configuration space for the general case of $n$-spheres in $n+1$-dimensional Euclidean space and indicate its conformal structure. This is followed by a discussion of a uniform distribution over the 4-dimensional de-Sitter configuration space used to model voids, that is $n=2$, in Section \ref{sec:uniform}. We shall derive, firstly, the corresponding void size distribution for voids in the whole (unrestricted) 3-space and find a self-similar power-law. Secondly, the effect of restricting void positions in 3-space due to survey geometry is studied, which naturally breaks the self-similarity of the size distribution. This is carried out analytically for a survey volume shaped as a spherical sector. Then in Section \ref{sec:comparison}, we investigate how the void size distributions predicted by the unrestricted and restricted uniform de-Sitter distributions compare with actual void size distributions. We consider, firstly, theoretical models of self-similarity and the excursion set formalism, to find that the power-law from the unrestricted uniform distribution is, to an extent, consistent with the expectation from structure formation theory. To illustrate this and the effect of survey geometry restriction in a data set, we apply an N-body simulation using the GADGET2 (\cite{Sp05}) code and extract spherical voids. A comparison with the fitting formula of \cite{vBeMu08} is given. We also discuss how the scaling dimension of the large scale structure can be estimated, using a fractal model of random spherical cutouts which is reviewed in the appendix. Section 5 contains our conclusions and comments on possible further applications of this idea.

\section{De-Sitter configuration space}
\label{sec:desitter}
We begin by briefly recapitulating the notion of a configuration space of spheres in Euclidean space and its de-Sitter geometry, as discussed in some more detail in \cite{GiWe13}; see also \cite{Ze13}, pp. 646--647. Consider an $n$-dimensional unoriented sphere $\mathbb{S}^n(R,\mathbf{x})$ in $n+1$-dimensional Euclidean space $\mathbb{E}^{n+1}$ given in terms of its radius $R>0$ and its centre at $\mathbf{x}=(x^1,\ldots x^{n+1})\in \mathbb{E}^{n+1}$. It turns out that, to each such sphere, one can uniquely assign a point $X^\mu=(X^0,\mathbf{X},X^{n+2}), \ 0\leq \mu \leq n+2$, in $n+3$-dimensional Minkowski space $\mathbb{E}^{1,n+2}$ endowed with the usual metric $\eta_{\mu\nu}=\mathrm{diag}(-1,1,\ldots,1)$, with 
\begin{eqnarray*}
X^0&=&-\frac{1}{2}\left(\frac{\mathbf{x}^2-1}{R}-R\right),\\
\mathbf{X}&=&-\frac{\mathbf{x}}{R},\\
X^{n+2}&=&-\frac{1}{2}\left(\frac{\mathbf{x}^2+1}{R}-R\right), 
\end{eqnarray*}
on the $n+2$-dimensional hypersurface 
\[
\eta_{\mu\nu}X^\mu X^\nu=1,
\]
which is a de-Sitter quadric. In other words, each sphere with a given centre and radius corresponds to a point on an $n+2$-dimensional de-Sitter space $dS^{n+2}$ which can therefore be interpreted as a configuration space of spheres. Since for unoriented spheres one may set $R> 0$,  we have  $X^0 - X^{n+2} = \frac{1}{R} >0$. Therefore, the space of unoriented spheres is in fact given by half of the full  de-Sitter quadric, sometimes referred to as de-Sitter space modulo the antipodal map $X^\mu \mapsto -X^\mu$. Geometrically, the coordinates $X^\mu$ of a sphere can be regarded as a form of Lie cycle coordinates for the Laguerre cycle representing the sphere (see, e.g., the appendix of \cite{GiWe13} for more mathematical details). Taking $y^i=(R,\mathbf{x}), \ 0\leq i \leq n+1$, as coordinates of the de-Sitter configuration space, its metric $g$ induced by the ambient Minkowski metric in the usual way can be read off from the line element
\begin{eqnarray}
\rmd s^2&=&\left(-(\rmd X^0)^2+ (\rmd X)^2+(\rmd X^{n+2})^2\right)|_{dS^{n+2}} \nonumber \\
\mbox{}&=&\frac{1}{R^2}\left(-\rmd R^2+\rmd \mathbf{x}^2\right) \nonumber \\
\mbox{}&=& g_{ij}\rmd y^i\rmd y^j
\label{lineelement}
\end{eqnarray}
so that, with respect to the coordinate-induced basis,
\begin{equation}
g_{ij}=\mathrm{diag}\left(-\frac{1}{R^2},\frac{1}{R^2},\ldots,\frac{1}{R^2}\right).
\label{ds-metric}
\end{equation}
\begin{figure}
\centering
\includegraphics[width=0.8\columnwidth]{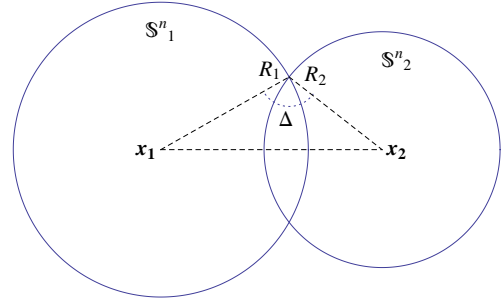} 
\caption{Angle between spheres. The intersection of two $n$-spheres $\mathbb{S}^n_1(R_1,\mathbf{x}_1),\mathbb{S}^n_1(R_1,\mathbf{x}_1)$ in $\mathbb{E}^{n+1}$, here illustrated for the case $n=1$ and, in projection, $n=2$, can be characterized by the Euclidean angle $\Delta(\mathbb{S}^n_1, \mathbb{S}^n_2)$ given in eq. (\ref{angle}), which is an invariant of the isometries of the de-Sitter configuration space.}
\label{fig:angle} 
\end{figure}
Hence, this metric measures the distance between spheres in terms of their configuration, that is, their radii and positions in space. Isometries of this de-Sitter space can be identified with conformal transformations of spheres, for the following reason. Consider two intersecting spheres $\mathbb{S}^n_1(R_1,\mathbf{x}_1)$ and $\mathbb{S}^n_2(R_2,\mathbf{x}_2)$ given by $X^\mu_1$ and $X^\mu_2$, respectively. Then
\begin{eqnarray}
\left(\eta_{\mu\nu}X^\mu_1 X^\nu_2\right)|_{dS^{n+2}}&=&\frac{R_1^2+R_2^2-(\mathbf{x}_1-\mathbf{x}_2)^2}{2R_1R_2}\nonumber \\
\mbox{}&=&\cos \Delta(\mathbb{S}^n_1,\mathbb{S}^n_2), 
\label{angle}
\end{eqnarray}
so that isometries leave the Euclidean angle $\Delta(\mathbb{S}^n_1, \mathbb{S}^n_2)$ between the two spheres invariant which, as illustrated in Fig. \ref{fig:angle}, measures their contact. Indeed, it provides a geometrical interpretation for the line element (\ref{lineelement}), since for two spheres that are infinitesimally close in the configuration space so that they are nearly identical in their radii and positions, one finds,
\begin{eqnarray*}
\rmd s^2&=&\frac{1}{R^2}(-\rmd R^2+\rmd \mathbf{x}^2)\\
\mbox{}&=&\rmd \Delta^2(\mathbb{S}^n(R,\mathbf{x}),\mathbb{S}^n(R+\rmd R,\mathbf{x}+\rmd \mathbf{x})).
\end{eqnarray*}
Finally, a measure of the number of spheres having radii within $[R,R+\rmd R]$ and centres in the volume element $\rmd x^1\ldots \rmd x^{n+1}$ at $\mathbf{x}\in\mathbb{E}^{n+1}$ is provided by the volume element,
\begin{eqnarray}
\rmd V^{n+2}&=& \sqrt{-\det g_{ij}} \rmd y^0 \ldots \rmd y^{n+1}\nonumber \\
\mbox{}&=&\frac{1}{R^{n+2}}\rmd R \rmd x^1 \ldots \rmd x^{n+1},
\label{ds-volume}
\end{eqnarray}
of the $n+2$-dimensional de-Sitter configuration space. 

\section{Uniform distribution over $\lowercase{d}S^4$}
\label{sec:uniform}
\subsection{Without survey geometry restriction}
\label{sec:unrestricted}
In order to study the distribution of cosmic voids from the geometrical perspective of the de-Sitter configuration space, we shall model them as unoriented $2$-spheres in $\mathbb{E}^3$. Hence specializing the previous discussion to the case $n=2$, the corresponding 4-dimensional configuration space has the de-Sitter geometry $dS^4$ with metric given by (\ref{ds-metric}). Then the configuration space volume element for voids with radii in $[R,R+\rmd R]$ and with centres in the volume element $\rmd x^1 \rmd x^2 \rmd x^3$ at $\mathbf{x} \in \mathbb{E}^3$ becomes
\begin{equation}
\rmd V^4=\frac{1}{R^4}\rmd R \rmd x^1 \rmd x^2 \rmd x^3, 
\label{ds-volume2}
\end{equation}
from eq. (\ref{ds-volume}). Also, the infinitesimal number of voids $\rmd N$ with radii in $[R,R+\rmd R]$ centred within the volume element at $\mathbf{x}$ can be described in terms of a non-negative distribution function over the configuration space, $f:dS^4\rightarrow \mathbb{R}^+$, so that
\begin{equation}
\rmd N=f(R,\mathbf{x}) \rmd R \rmd x^1 \rmd x^2 \rmd x^3. 
\label{f1}
\end{equation}
In this section, we will consider the simplest distribution in this framework, corresponding to a uniform distribution of voids over the de-Sitter configuration space such that
\[
\rmd N \propto \rmd V^4.
\]
Clearly, this uniform distribution has the property that all voids are counted regardless of their position in $\mathbb{E}^3$: there is no restriction due to survey geometry and any mutual overlap of voids is allowed. The corresponding distribution function $f$ is independent of the position in space, 
\begin{equation}
f(R, \mathbf{x}) \propto \frac{1}{R^4},  
\label{f2}
\end{equation}
from eq. (\ref{ds-volume2}) and (\ref{f1}). Defining the void number density as the number of void centres per volume element in $\mathbb{E}^3$, one can compute the cumulative number density $n(>R)$ of voids with radius greater than $R$ that corresponds to the uniform distribution,
\[
n(>R)=\int_R^\infty f(R') \rmd R' \propto \frac{1}{R^3},
\]
by integrating (\ref{f1}) using (\ref{f2}). Expressing this in terms of void volume $V=4\pi R^3/3$,
\begin{equation}
n(>V)= \frac{C}{V},
\label{cumul}
\end{equation}
where $C>0$ is a constant whose geometrical interpretation we shall return to in Section \ref{sec:self-similarity}. A more realistic modification of this cumulative number density should take into account that any actual void survey can, of course, only encompass a finite subregion of $\mathbb{E}^3$. We shall investigate this aspect in the following.   

\subsection{With survey geometry restriction}
\label{sec:restricted}
\begin{figure}
\centering
\includegraphics[width=0.8\columnwidth]{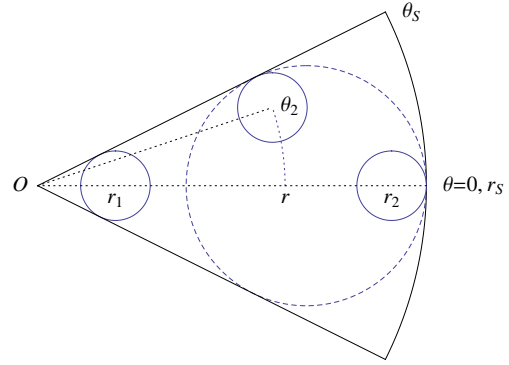} 
\caption{Survey geometry. In a survey region shaped as spherical sector with $r\leq r_S,\ \theta\leq \theta_S$ centred at the observer $O$, the limits $r_1(R),r_2(R)$ and $\theta_2(r,R)$ for voids of radius $R$ (solid circles) are illustrated, see eq.s (\ref{limits}). The largest void for the given survey geometry, at $r_0$ with radius $R_0$ given by eq. (\ref{largest}), is also shown (dashed circle).}
\label{fig:survey} 
\end{figure}
\begin{figure}
\centering
\includegraphics[width=\columnwidth]{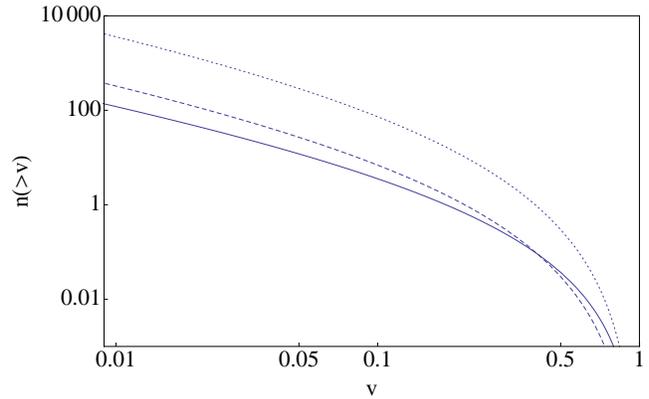} 
\caption{Cumulative number density of voids. Double logarithmic plots of $n(>v)$, the integral (\ref{cumul2}) with the same constant of proportionality and $r_S$ set to unity, as functions of the dimensionless volume parameter $0\leq v\leq 1$ defined in (\ref{v}), for three choices of the opening angle $\theta_S$ of the survey geometry: a solid curve for $\theta_S=\pi/2$, the half sphere; a dashed curve for $\theta_S=\pi/6$; and a dotted curve for $\theta_S=\pi/18$.}
\label{fig:cumul} 
\end{figure}
The assumption of a uniform distribution of voids over their de-Sitter configuration space results in a differential number density
\begin{equation}
\frac{\rmd n}{\rmd R}=f(R) \propto \frac{1}{R^4} \quad \Rightarrow \quad \frac{\rmd n}{\rmd V}=\frac{\rmd R}{\rmd V}\frac{\rmd n}{\rmd R}\propto \frac{1}{V^2}
\label{density}
\end{equation}
of  $\rmd n$ void centres per volume in $\mathbb{E}^3$, per void radius $\rmd R$ or void volume $\rmd V$, respectively, again by eq. (\ref{f1}) and (\ref{f2}). Now suppose that a survey counts only those voids which are wholly within the survey region $V_S$. For simplicity, consider a survey geometry shaped as a spherical sector with opening angle $\theta_S$ and maximal (comoving) distance $r_S$ from an observer who is situated at its apex and coordinate origin. Then, in terms of spherical polar coordinates, the survey region can be described by
\[
V_S: \quad 0\leq r \leq r_S, \quad 0\leq \theta \leq \theta_S, \quad 0 \leq \phi \leq 2\pi, 
\]
and we let $\theta_S\leq \pi/2$. The total volume of the survey is then
\[
V_S(r_S,\theta_S)=\frac{2\pi}{3}r_S^3(1-\cos \theta_S).
\]
A given void of radius $R$ will be counted in the distribution if it lies wholly within $V_S$, that is, if the coordinates of its centre $(r,\theta,\phi)$ satisfy the conditions
\[
V_S(R): \ r_1(R)\leq r \leq r_2(R), \ 0 \leq \theta \leq \theta_2(r,R), \ 0\leq \phi \leq 2\pi,
\]
where the lower and upper coordinate limits are determined by voids touching the boundary of $V_S$. The conditions for these can be read off from Fig. \ref{fig:survey},
\[
r_1(R)\sin \theta_S=R, \ r_2(R)+R=r_S, \ r\sin(\theta_S-\theta_2(r,R))=R,
\]
whence we have
\begin{eqnarray}
r_1(R)&=& \frac{R}{\sin\theta_S}, \nonumber \\
r_2(R)&=&r_S-R,\label{limits} \\
\theta_2(r,R)&=&\theta_S-\arcsin\frac{R}{r}. \nonumber
\end{eqnarray}
The largest void which fits into $V_S$ is situated at $r_0$ on the polar axis and has radius $R_0$, whose values can be obtained from the condition $r_1(R_0)=r_2(R_0)$. 
Hence, using eq. (\ref{limits}),
\begin{equation}
r_0=\frac{r_S}{1+\sin\theta_S}, \quad R_0=\frac{r_S \sin \theta_S}{1+\sin \theta_S}.
\label{largest}
\end{equation}
In order to compare different survey geometries, it will be convenient to express void volume $V$ in terms of the volume $V_0$ of the largest void admitted by a given survey geometry $(r_S,\theta_S)$, and hence define the dimensionless quantity $0\leq v \leq 1$,
\begin{equation}
v=\frac{V}{V_0(r_S,\theta_S)} \quad \mbox{such that} \quad R=\frac{r_S \sin \theta_S}{1+\sin \theta_S}v^\frac{1}{3}  
\label{v}
\end{equation}
from eq. (\ref{largest}). The differential number density of voids within the range $[v,v+\rmd v]$ of the volume parameter can then be expressed as
\[
\rmd n=\frac{1}{V_S(r_S,\theta_S)}\int_{V_S(v)}\rmd N, 
\]
so that by eq. (\ref{f1}) in spherical polar coordinates,
\begin{equation}
\frac{\rmd n}{\rmd v}=\frac{1}{V_S(r_S,\theta_S)}\frac{\rmd R}{\rmd v}\int_{V_S(v)}f(v,r,\theta,\phi)r^2 \sin \theta \rmd r \rmd \theta \rmd \phi. 
\label{density1}
\end{equation}
Using the limits of $V_S(v)$ for voids with volume parameter $v$ given by (\ref{limits}) and the uniform distribution (\ref{f2}), whose constant of proportionality is kept, we can recast (\ref{density1}) to obtain
\begin{eqnarray}
\frac{\rmd n}{\rmd v}&\propto&\frac{(1+\sin\theta_S)^3}{\sin^3\theta_S(1-\cos\theta_S)r_S^6}\frac{1}{v^2}\int_{r_1}^{r_2}\int_0^{\theta_2} r^2\sin\theta \rmd \theta \rmd r \nonumber \\
\mbox{}&=&\frac{(1+\sin\theta_S)^3}{3\sin^3\theta_S(1-\cos\theta_S)r_S^3}\frac{1}{v^2} \left[ 1\phantom{\int}\right. \nonumber \\
\mbox{}&\phantom{=}&-\frac{3}{2}\frac{\sin\theta_S(2+\sin\theta_S)}{1+\sin\theta_S}v^\frac{1}{3}+\frac{\sin^2\theta_S}{1+\sin\theta_S}\left(3v^\frac{2}{3}-\frac{v}{2}\right)\nonumber\\
\mbox{}&\phantom{=}&-\left.\cos\theta_S\left(1-\frac{2\sin\theta_S}{1+\sin\theta_S}v^\frac{1}{3}\right)^\frac{3}{2}\right] \label{density2}.
\end{eqnarray}
This expression can in turn be integrated to yield the corresponding cumulative number density,
\begin{equation}
 n(>v)=\int_v^1 \frac{\rmd n}{\rmd v'}\rmd v',
\label{cumul2}
\end{equation}
and Fig. \ref{fig:cumul} shows it for three choices of the opening angle $\theta_S$ of the survey geometry. These results for the number density (\ref{density2}) and cumulative number density of voids illustrate how the intrinsic distribution and the observed distribution can differ substantially due to the restrictions imposed by the survey geometry. Small voids, of course, tend to be affected less by the survey geometry and therefore approach asymptotically the power-law (\ref{cumul}) of the intrinsic distribution.  

\section{Comparison with void distributions}
\label{sec:comparison}
\subsection{From theoretical models}
\subsubsection{Self-similarity}
\label{sec:self-similarity}
As shown in Section \ref{sec:unrestricted}, the uniform de-Sitter distribution without survey geometry restriction yields a power-law size distribution given by 
\begin{equation}
\nu(V):=\frac{\rmd n}{\rmd V}\propto \frac{1}{V^2}
\label{square}
\end{equation}
for the differential number density of voids, from eq. (\ref{density}). This is obviously self-similar in the sense that a change in volume scale, $\tilde{V}=sV, \ s=\mathrm{const.}$, also implies 
$\nu(\tilde{V})\propto \tilde{V}^{-2}$. Interpreted as a probability distribution, we can identify this as a Pareto distribution of index one. Considering, on the other hand, a sufficiently large but finite set of voids in a fixed volume of $\mathbb{E}^3$, then the integer rank $\mathcal{R}(V)$ of a void of volume $V$ in a decreasing ordered sequence of volumes is proportional to $n(>V)\propto V^{-1}$. Hence, one can see that Zipf's law is satisfied,
\[
V\propto\frac{1}{\mathcal{R}(V)}.
\]
Self-similarity and power-laws are a characteristic feature of fractals, and their applicability to cosmic voids has been investigated early on (e.g., \cite{Ma82}). More recently, Zipf's law in this context has been studied by \cite{GaMa02}, followed by more general considerations of the fractal properties of void distributions (e.g., \cite{Ga06,Ga07,Ga09}) including the notion of multifractals, which admits a range rather than a single fractal dimension, and lacunarity, which characterizes the void distribution at a given fractal dimension. This is, of course, also related to the question whether and to what extent the large scale structure itself can be described by fractals (see, e.g., the review by \cite{Jo-etal04}). Denoting the (comoving) distances of galaxies in $\mathbb{E}^3$ by $x=|\mathbf{x}|$, one can define a local scaling index in terms of the cumulative number count of galaxies $N_{\mathrm{gal}}(<x)$  (c.f. \cite{Jo-etal04}, pp. 1244--1245),
\[
D(x,x_0)=\frac{\log  \frac{N_{\mathrm{gal}}(<x)}{N_{\mathrm{gal}}(<x_0)} }{\log \frac{x}{x_0}}.
\]
This can be interpreted as a (single) fractal dimension if it is constant and independent of the scale $x_0$, for instance the Hausdorff dimension of a Cantor set if $0<D<2$ (cf. Appendix A), in which case the corresponding number count is simply a power-law of the form $N_{\mathrm{gal}}(<x)\propto x^D$. If the homogeneity postulate of the cosmological principle holds, one expects $D\rightarrow 3$ as $x\rightarrow \infty$. The detailed interpretation of $D$, however, depends on the homogeneity scale $x_{\mathrm{hom}}$, the correlation length $x_{\mathrm{cor}}$ and sample size $x_{\mathrm{max}}$ (c.f. \cite{Ga-etal05}, pp. 236--240), as well as on the finiteness of the sample number (e.g., \cite{BaYaSe08}).
\newline
It is interesting to note that a simple fractal model, Mandelbrot's \textit{random tremas} interpreted as voids, provides another geometrical perspective on the cumulative number density of eq. (\ref{cumul}) implied by the uniform de-Sitter distribution, which also connects its constant of proportionality $C$ with the fractal dimension $D$ of the large scale structure mentioned above. Consider $\mathbb{E}^3$ with a homogeneous mass density and introduce a sequence of spherical cutouts (tremas) to model voids. These spheres are placed randomly so that arbitrary overlap is allowed. For a particular size distribution of these spheres, it turns out that an infinite sequence will not empty the space completely but leave in general a non-empty complementary set of some Hausdorff dimension $D$. The cumulative number density of the spheres in this case, counting all regardless of overlap, is (cf. Appendix A)
\begin{equation}
n(>V)=\left(1-\frac{D}{3}\right)\frac{1}{V},
\label{trema1}
\end{equation}
which has the same scaling behaviour as the unrestricted uniform de-Sitter distribution, for which arbitrary overlap of spheres is also allowed as discussed in Section \ref{sec:unrestricted}. Interpreting the unrestricted uniform de-Sitter distribution in terms of the random trema model, we can therefore identify 
\begin{equation}
C=1-\frac{D}{3}
\label{trema2}
\end{equation}
from eq. (\ref{cumul}). In the case of a finite data set, one should require that the sample size is much greater than the correlation length, $x_{\mathrm{max}} \gg x_{\mathrm{cor}}$, for eq. (\ref{trema2}) to apply, as will be discussed further in Section \ref{sec:discussion} below. Finally, it may be emphasized that it is the overlap of the cutouts allowed in this model which causes the fixed exponent in the power-law $n(>V)\propto V^{-1}$ and relegates information about $D$ to the constant of proportionality. This is different from the disjoint cutouts considered more usually, e.g. by \cite{Ga06}, which contain $D$ in the exponent.

\subsubsection{Excursion set formalism}
\label{sec:excursion}
In their pioneering paper, \cite{PrSc74} point out that, starting from Gaussian density perturbations in a Friedmann cosmology and considering the linear quasi-Newtonian perturbation analysis of the growing mode, the mass distribution at late times does not depend on the initial mass distribution. They identify a simple physical reason for this self-similarity in the existence of two dimensionless quantities governing the gravitational collapse, which remain approximately \textit{constant} during the matter-dominated phase on subhorizon scales. The Press-Schechter argument yielding a power-law size distribution can be summarized as follows. Consider massive particles distributed in some comoving volume. Then the mass density function $\rho$ on this volume is obtained by applying some smoothing window function with scale length $l$ at each point. At sufficiently early times, this density function will be approximately constant,
\[
\rho=\bar{\rho}(1+\delta), \quad \left|\delta\right|\ll 1. 
\]
However, if the density within a window is greater than some (fixed dimensionless, say) critical density so that $\delta>\delta_{\mathrm{crit}}$ around overdensity peaks due to fluctuations in the density function, then the mass within the window will gravitationally collapse and ultimately form a bound object. Again, for sufficiently early times, one may take these fluctuations to be Gaussian, such that the volume fraction of points with collapsing windows is given by
\[
\mathcal{F} \propto \frac{1}{\sigma}\int_{\delta_\mathrm{crit}}^\infty \exp \left(-\frac{\delta^2}{2\sigma^2}\right)\rmd \delta.
\]
These windows will contain slightly different masses, but basically $M\propto \bar{\rho} l^3$. Also, the variance may depend on the smoothing scale and hence also on the mass, $\sigma^2\propto l^{-k}\propto M^{-k/3}$, for some spectral index $k\geq0$. Thus, $\mathcal{F}$ is a function $M$ as well, and so the differential number density $\nu(M)$ of ultimately bound objects obeys 
\begin{equation}
M\nu(M)\propto -\bar{\rho}\frac{\rmd \mathcal{F}}{\rmd M}=-\bar{\rho} \frac{\rmd \sigma}{\rmd M}\frac{\rmd \mathcal{F}}{\rmd \sigma},
\label{ps2}
\end{equation}
and hence
\[
\nu(M)\propto \left(\frac{M}{M_0}\right)^{\frac{k}{6}-2}\exp \left[-\left(\frac{M}{M_0}\right)^{\frac{k}{3}}\right],
\]
where $M_0$ is the mass scale of the exponential term, which gives rise to a cutoff at large masses (or volumes). For $M \ll M_0$ and a scale-invariant $\sigma$ with $k=0$, this yields the inverse square power-law
\begin{equation}
\nu(M)\propto M^{-2}.
\label{pslimit}
\end{equation}
It turns out that this power-law emerges also in the modified Press-Schechter approach proposed by \cite{ApJo90}, which uses an adaptive window scale $l$. Now in the context of voids, one can consider underdensity troughs rather than overdensity peaks (as in \cite{Ba-etal86}) and, since Gaussian fluctuations are symmetric about the average density, the same power-law applies, as pointed out by \cite{ShvdWe04}. Then since $M\propto l^3\propto V$, we recover (\ref{square}) from (\ref{pslimit}). Of course, in addition to the large volume cutoff, this simple excursion set argument ignores effects of void hierarchy which affect small volumes in particular (i.e., by the void-in-void and void-in-cloud processes, cf. \cite{ShvdWe04}), so that the power-law will only apply within a range of volumes. 
\newline
Hence, at least within a range of volumes, the power-law (\ref{square}) has a physical basis in the excursion set formalism: the Gaussian fluctuations producing voids give rise to a self-similar power-law size distribution which can be described geometrically as a uniform distribution over the de-Sitter configuration space of spheres.

\subsection{From data}
\label{sec:voidfinder}
\subsubsection{An N-body simulation}
Now in order to compare the void size distributions predicted by the unrestricted and restricted uniform de-Sitter distributions from Section \ref{sec:uniform} with data, we perform an N-body simulation with the parallel Tree-Particle Mesh code, GADGET2 (\cite{Sp05}),  using cosmological parameters consistent with the WMAP seven-year results (\cite{Ko-etal11}), namely spatial curvature $k=0$, dark energy density $\Omega _\Lambda = 0.7274$, Hubble parameter $h=0.704$ and spectral index $n_s=0.963$. The initial condition for the simulation is generated at redshift $z_0=50$ using the code developed by \cite{Ni-etal09}, who employ second order Lagrangian perturbation theory. The simulation is run in a cubic box with side length $240h^{-1}$Mpc in comoving coordinates with periodic boundary conditions, and $256^3$ dark matter particles whose individual mass is scaled to match the mass density of the Universe. At redshift $z=0$, voids are extracted according to the following algorithm.
\newline
Firstly, the discrete simulation data are smoothed to construct the matter density function, using a Gaussian window with adaptive length scale. This smoothing length is chosen to be the distance to the 20th nearest particle at each cell. Secondly, a set of spherical voids is extracted by defining density minima as void centres and determining the radius for each void as the largest radius $R$ centered at the density minimum for which the average density of the enclosed sphere $\rho(R)$ is less than a critical value
\[
\delta _{\mathrm{crit}} =\frac{\rho(R)-\bar{\rho}}{\bar{\rho}}= -0.8,
\]
relative to the average density $\bar{\rho}$ of the box. Hence, for our present purposes, we allow arbitrary overlap of the spherical voids and do not apply an additional merging algorithm to these voids (cf. \cite{Co-etal05}, also called \textit{protovoids}), which would give rise to non-spherical voids. This should allow for are meaningful comparison with the uniform distribution over the de-Sitter configuration space of spheres, which allows void overlap as noted in Section \ref{sec:unrestricted}.

\subsubsection{Discussion}
\label{sec:discussion}
\begin{figure}
\vspace*{-4mm}
\hspace*{-5mm}
\rotatebox{-90}{
\includegraphics[width=0.76\columnwidth]{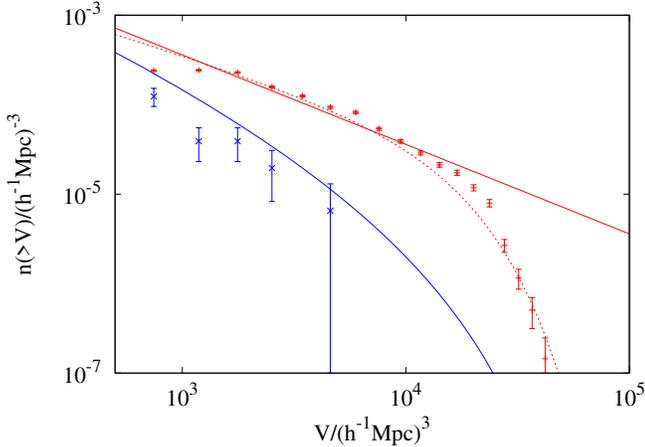}}
\caption{Simulated void size distributions. Double logarithmic plots of the cumulative number density of voids as a function of volume, using the simulation discussed in Section \ref{sec:voidfinder}: uniform de-Sitter distribution from eq. (\ref{cumul}), $C=0.36$, without survey geometry restriction (upper solid curve and data set) and fitting formula of von Benda-Beckmann \& M\"{u}ller (2008) (dotted); uniform de-Sitter distribution from eq. (\ref{cumul2}) with survey geometry restriction in spherical sector (lower solid curve and data set). Poisson errors are indicated.}
\label{fig:plot} 
\end{figure}
The resulting cumulative number density of voids in the simulation box as a function of void volume is shown in Fig. \ref{fig:plot}. The largest void has a volume of $V_{\mathrm{max}}\simeq 4.2\cdot 10^4 (h^{-1}$Mpc$)^3$, which can be interpreted in terms of the exponential cutoff in the excursion set formalism mentioned in Section \ref{sec:excursion}, and whose value is comparable with the simulation results reported in Fig. 7 of \cite{Co-etal05}. Until the curve flattens at very small volumes -- here, of course, rather due to simulation resolution and smoothing scale than effects of void hierarchy --, the cumulative number density for $V\ll V_{\mathrm{max}}$ can indeed be approximated by the power-law of eq. (\ref{cumul}) derived from the unrestricted uniform de-Sitter distribution. It is also instructive to compare this result with the fitting formula of \cite{vBeMu08}, which was shown to model well the cumulative volume filling factor $F(>R)$ of voids derived from magnitude-limited samples of galaxies in the 2dF Galaxy Redshift Survey,
\begin{equation}
F(>R) = \exp \left [-\left (\frac{R}{s_1\lambda }\right )^{p_1}-\left (\frac{R}{s_2\lambda }\right )^{p_2}\right ],
\label{fitting}
\end{equation}
where $R=(3V/4\pi)^{1/3}$ is the effective spherical radius of a void of volume $V$ and any shape, $\lambda$ denotes the mean separation between galaxies, and $p_1,p_2,s_1,s_2$ are parameters. In order to convert this to a cumulative number density as a function of volume, one needs to rewrite eq. (\ref{fitting}) in terms of $V$. Introducing a factor $c>1$ to correct for the fact that our voids are not merged and therefore have a higher number density, we let
\[
F(>V)=\frac{1}{c}\int_V^\infty V'\nu(\rmd V')\rmd V', 
\]
where $\nu$ denotes the differential number density as before, to obtain by differentiating
\[
\nu(V)=-\frac{c}{V}\frac{\rmd F(>V)}{\rmd V}, 
\]
so that
\[
n(>V)=\int_V^\infty\nu(\rmd V')\rmd V'=-\int_V^\infty\frac{c}{V'}\frac{\rmd F(>V')}{\rmd V'} \rmd V' 
\]
can be computed from (\ref{fitting}). While a detailed comparison is beyond the scope of this paper, we note that the parameters $p_1=1$, $p_2=4.5$, $s_1=1$, $s_2=1.5$ and $\lambda=11 \ h^{-1}$Mpc, which seem appropriate choices given the data in Tables 1 and 2 of \cite{vBeMu08}, yield a curve in reasonable agreement with our simulation data, as seen in Fig. (\ref{fig:plot}), using a factor of $c=2.5$ which is compatible with the void number densities shown in Fig. 6 and Fig. 7 of \cite{Co-etal05}. We also note that the constant of proportionality in eq. (\ref{cumul}) which fits the power-law range is $C\simeq0.36$. Thus, as described in Section \ref{sec:self-similarity} , one would expect the fractal dimension of the large scale structure to be $D\simeq 1.9$ by eq. (\ref{trema2}), which seems reasonable given that real observational data can be approximated by $D\simeq 1.2$ at small scales transitioning to $D\simeq 3$ at large scales, with a homogeneity scale of $x_{\mathrm{hom}}\gtrsim 10 h^{-1}$Mpc and a correlation length of $x_{\mathrm{cor}} \simeq 6 h^{-1}$Mpc (e.g. \cite{Jo-etal04}, pp. 1231--1235, Table 1, and references therein), so that our box size is indeed much larger than the correlation length.
\newline
Given, then, that our simulation appears to produce a realistic size distribution of spherical voids without merging, which does indeed agree with the uniform de-Sitter distribution for voids in the range $10^3\ldots10^4 (h^{-1}$Mpc$)^3$, we shall now turn to the effect of restricting the survey geometry. To this end, we select a survey region in the simulation box shaped like a spherical sector as in Fig. \ref{fig:survey}, with its apex at a corner of the box and its axis oriented along the diagonal. Consider such a spherical sector with an opening angle of $\theta_S=10^\circ$ and a radius of half the maximum radius within the box, $r_S\simeq 169 h^{-1}$Mpc. Although such a survey volume has only $1.1\%$ of the total box volume, the volume of the largest void within this survey region is $V_0\simeq 6.5\cdot 10^4 (h^{-1}$Mpc$)^3$, from eq. (\ref{largest}), which is greater than $V_{\mathrm{max}}$. The corresponding cumulative number density is also shown in Fig. \ref{fig:plot}, together with the theoretical prediction of eq. (\ref{cumul2}) from the restricted uniform de-Sitter distribution. Since, for voids in the whole box, the unrestricted uniform de-Sitter distribution is a good approximation in the range $10^3\ldots10^4 (h^{-1}$Mpc$)^3$ where $v=V/V_0$ is of the order $0.1$, we expect from Fig. \ref{fig:cumul} that the survey geometry restriction causes a significant deviation in this range from the power-law asymptote. Furthermore, since $V_0/V_{\mathrm{max}}\simeq 1.5$, we expect a small effect of void overcounting in the restricted uniform distribution due to the large volume cutoff. Both expectations are indeed borne out by the data, as can be seen in the figure.

\section{Concluding remarks}
\label{sec:conclusions}
\begin{quote}
\textit{``In parallel with efforts to explain, I think it indispensable to describe clustering, and to mimic reality by purely geometric means."}
\newline
\cite{Ma82}, p. 84
\end{quote}
In this article, we have considered an application of the 4-dimensional de-Sitter configuration space of 2-spheres in Euclidean 3-space to cosmology. Modelling cosmic voids as spheres and allowing any overlap, it was shown that a uniform distribution over this configuration space gives rise to a self-similar power-law size distribution of voids. It appears to agree well with data in an intermediate range of void volumes, and this can be understood physically from the excursion set formalism as long as the large volume cutoff and void hierarchy effects at small volumes may be ignored. Moreover, we pointed out that this power-law may also be reinterpreted in terms of Mandelbrot's random trema model, allowing one to estimate the fractal dimension of the large scale structure in a different way compared to other approaches. We have also seen how restrictions of void positions in 3-space due to survey geometry significantly affect the size distribution.
\newline
Now in order to refine this approach, one might implement void merging conditions using this configuration space language. Furthermore, since the uniform distribution appears to provide a reasonably good match with actual void distributions of intermediate volume, it may be worthwhile to treat the uniform distribution as a null hypothesis and study the physical interpretation of non-uniform distributions. Then deviations from uniformity will, as we have seen, encode physically interesting effects such as void hierarchy processes or the large volume cutoff, which may be particularly sensitive to the underlying cosmology.     
\newline
Finally, while de-Sitter geometry plays an important and well-known r\^{o}le in cosmological spacetimes, we would like to emphasize the interesting, and perhaps rather surprising, fact that the present application of de-Sitter geometry to cosmology arises in a physically entirely different context. At this stage, our approach is primarily a descriptive device which also offers a novel geometrical interpretation of an aspect of self-similarity in cosmology, in the spirit of the quotation above.

\section*{Acknowledgments}
MCW gratefully acknowledges support from the World Premier International Research Center Initiative (WPI Initiative), MEXT, Japan, and would like to thank Masahiro Takada and Jean-Philippe Uzan for useful discussions. NY acknowledges financial support from the Japan Society for the Promotion of Science (JSPS) Grant-in-Aid for Scientific Research (25287050).

\section*{Appendix A: Random trema fractals}
Here we give a simple heuristic argument for the cumulative number density of the random trema model of Section \ref{sec:self-similarity}, as stated in eq. \ref{trema1}. This idea was introduced by Mandelbrot and is sketched in \cite{Ma82}, pp. 281--283 and pp. 301--302. See also \cite {Ga-etal05}, pp. 372--373. A rigorous exposition of the present argument can be found in \cite{Fa97}, pp. 136--142, where the relevant mathematical literature is also cited. The notion of disjoint cutouts as used, e.g., by \cite{Ga06} and discussed, e.g., in \cite{Fa97}, pp. 51--56, is different from the one reviewed here.
\newline
Consider, in $\mathbb{E}^1$ for simplicity, the closed interval $[0,L_0]$ of the real line, from which an infinite sequence of open subintervals with decreasing length $L_i$, where $0<L_i<\frac{1}{2}$ and $i \geq 1$, are removed, which Mandelbrot calls \textit{tremas} and which are used to model voids. These subintervals are placed randomly and may overlap, that is, their midpoints are distributed uniformly and independently over the interval $[0,L_0]$ whose endpoints are identified. If the sum of subinterval lengths counted regardless of overlap diverges,
\[
\sum_{i=1}^k L_i \rightarrow \infty \quad \mbox{as} \quad k\rightarrow \infty,
\]
it is plausible that the overlapping subintervals will almost certainly remove the entire interval $[0,L_0]$, depending on how fast this sum diverges. Now it turns out that if
\[
L_i = \frac{CL_0}{i},
\]
where $0<C<\frac{1}{2}$ is a given constant number, so that the sum of subinterval lengths diverges only slowly as a harmonic series,
\[
\sum_{i=1}^k L_i \rightarrow CL_0 \ln k \quad \mbox{as} \quad k \rightarrow \infty,
\]
then, with positive probability, the set remaining of the interval $[0,L_0]$ is in fact non-empty with Hausdorff dimension $D=1-C$ (cf. proposition 8.8, p. 140, in \cite{Fa97}). While the rigorous argument is technical, some insight into the origin of this result can be gained by estimating the Hausdorff dimension with a box-counting algorithm in an early stage with negligible overlap. Thus, at the $i$th stage of the sequence, the remaining length of the interval will be
\[
L(i)\simeq L_0(1-C \ln i) \quad  \mbox{with}  \quad C \ln i \ll1.
\]
Thinking of the sequence as dividing the interval successively into intervals of length $L_0/i$ at stage $i$, the corresponding number of intervals is
\[
N(i)=\frac{iL(i)}{L_0}\simeq i(1-C\ln i),
\]
and then the Hausdorff dimension can be computed by interval-counting,
\[
D=\frac{\log N(i)}{\log i}\simeq 1+\frac{\ln (1-C\ln i)}{\ln i} \simeq 1-C,
\]
as required.
\newline
Now the cumulative number density of these one-dimensional voids, counting all subintervals regardless of overlap, is 
\[
n(>L_i)=\frac{i-1}{L_0}=\frac{C}{L_i}-\frac{1}{L_0},
\]
so that for sufficiently large $i$ or small voids of length $L \ll CL_0$, we obtain a self-similar power-law  as a limiting case,
\[
n(>L)=(1-D)\frac{1}{L}.
\]
This can be generalized to $\mathbb{E}^{n+1}$ and a random trema model of $n$-spheres as voids. Then the cumulative number density of overlapping voids in terms of their $n+1$-volume $V$ becomes (cf. p. 302 in \cite{Ma82}),
\[
n(>V)=\left(1-\frac{D}{n+1}\right)\frac{1}{V},
\]
with $n=2$ in eq. (\ref{trema1}).
\newline
In addition to the Hausdorff dimension of the remaining set, which is a real number and a metric concept, one can also define the Lebesgue covering dimension $D_T$, which is an integer and a topological concept, and in general $D_T \leq D \leq n+1$. Now if $D$ exceeds $D_T$, the set is called \textit{fractal}. For a random set, it can be shown using an intersection argument that $D<n/2+1$ is a sufficient condition for $D_T=0$ (cf. \cite{Ma82}, pp. 215--216), which is called a Cantor set. Thus, we expect that a random trema model in 3-dimensional Euclidean space gives rise to a fractal Cantor set if $0<D<2$, as seems to be indicated by the data, c.f. Section \ref{sec:discussion}.
\label{lastpage}
\end{document}